\documentstyle[aps,prl,multicol,epsfig]{revtex}
\begin{document}
\title{ Investigation of trial wavefunction  approach to bilayer Quantum Hall systems }
\author{    Gun Sang Jeon and Jinwu Ye }
\address{ Department of Physics, The Pennsylvania State University, University Park, PA, 16802}
\date{\today}
\maketitle
\begin{abstract}
  We study properties of some known trial wavefunctions in bilayer quantum Hall
  systems at the total filling factor $ \nu_{T}=1 $. In particular, we find that
  the properties of a meron wavefunction and a natural "quasi-hole" wave
  function are dramatically different due to the broken symmetry and the associated Goldstone
  mode in the bulk. Although the ( smallest ) meron has localized charge $ 1/2 $
  and logarithmically divergent energy, the charge of the quasi-hole excitation
  extends over the whole system and its energy diverges linearly with the area of
  the system.  This indicates that the natural quasi-hole wavefunction is not a
  good trial wavefunction for excitations. It also shows that the energy of the
  naive candidate for a pair of meron wavefunction written down previously increases
  quadratically instead of logarithmically as their separation increases. Our results
  indicate that qualitatively good trial wave functions for the ground state and
  the excitations of the interlayer coherent bilayer quantum Hall system at finite
  $ d $ are still not available and searching for them remains an important open problem.
\end{abstract}

\section{Introduction}
   The variational approach has been successful for investigating condensed matter systems,
  especially in superconductivity and Quantum Hall Effects.
  In 1957, Bardenn, Cooper and Schrieffer proposed BCS trial wave-function for a $ S $-wave
  superconducting ground state and also
  constructed all possible quasi-particle excitations above the ground state.
  In 1983, Laughlin proposed a trial-wave function for the ground states of
  Fractional Quantum Hall Effects (FQHE) \cite{laug} at filling factors $ \nu=\frac{1}{2s+1} $.
  He also constructed trial wavefunctions of  quasi-hole and quasi-electron excitations and
  found that they are localized and carry fractional charges $ \pm \frac{1}{2s+1} $.
  Jain proposed Composite
  Fermion (CF) trial wavefunctions for the ground states at Jain's principal series at filling factors
  $ \nu= \frac{p}{ 2 s p \pm 1} $ \cite{jain}. CF theory  produces 
  a spectrum of the low energy CF excitations which matches
  that obtained from exact diagonalization study in finite size systems.
  These excitations are localized and carry fractional charges $ \frac{1}{ 2 s p \pm 1} $.
   Halperin generalized Laughlin's wavefunctions to
   $ (m, m^{\prime}, n ) $ wavefunctions  ( See Eqn.\ref{mn} ) for FQHE in a two-component system \cite{bert}.
   These components could be the spins of electrons
   when the Zeeman coupling is very small or layer indices in multi-layered system.
   Quasi-hole and quasi-particle wavefunctions for generic $ (m, m^{\prime}, n ) $ state
   in two-component systems can also be constructed by analogy to
   those in the single layer systems ( see Eqn. \ref{qhmn}). Scarola and Jain constructed a
   general class of CF wavefunctions in bilayer Quantum Hall systems which include Halperin's
   $ (m, m^{\prime} , n ) $ states as a subset and
   investigated the energies of these wavefunction
   as a function of the distance $ d $ between the two layers
   \cite{sca}.

   One of the most interesting cases is the
   spin-polarized Bilayer Quantum Hall systems (BLQH) at total filling factor
   $ \nu_{T} =1 $. For sufficiently small interlayer separation $ d $,
   Halperin's $ (111) $ wavefunction ( see Eqn.\ref{111} ) is believed to be a good trial
   wave-function for the ground state.
  In contrast to general $  (m, m^{\prime}, n ) $ state,
  the $ (111) $ state ( or $ ( m, m, m ) $ state in general ) is a symmetry broken state
  and there is an associated gapless Goldstone mode in the bulk \cite{fer,wen,jap}.
  This system has been a subject of enormous experimental and theoretical investigation over the last decade
  \cite{yang,moon,rev}.  When the interlayer separation $ d $ is sufficiently large, the bilayer
  system decouples into two separate compressible $ \nu=1/2 $ layers.
  However, when $ d $ is smaller than a critical distance $ d_{c} $, in the absence
  of interlayer tunneling, the system undergoes a quantum phase transition into
  a novel spontaneously interlayer coherent incompressible phase \cite{rev}.
  Recently, at low temperature, with extremely small interlayer tunneling amplitude,
  Spielman {\sl et al} discovered
  a very pronounced narrow zero bias peak in this interlayer coherent
  incompressible state \cite{gold}. M. Kellogg {\sl et al } also observed
  quantized Hall drag resistance at $ h/e^{2} $ \cite{hall}. 

   By treating the layer index as a pseudo-spin index,
   Girvin, Macdonald and collaborators mapped the bilayer system
   into a Easy Plane Quantum Ferromagnet (EPQFM) \cite{rev} and
   explored many rich and interesting physical phenomena in this system.
   The low energy physics is given by an effective $ 2+1 $ dimensional $ XY $ model.
   They interpreted the gapless mode as the spin wave mode of quantum XY model.
    and the vortex excitations of the XY model as topological defects called " merons "
   which carry fractional charges $ \pm 1/2 $ and
   also have $ \pm $ vorticities. The  merons have logarithmically divergent self energies and are bound in pairs
   at low temperature. The lowest energy excitations carry charge $ \pm e $ which
   are a meron pair with opposite vorticity.
   By drawing analogy with superconductivity, they wrote down the EPQFM wavefunction Eqn.\ref{ex} 
   which can also be interpreted
   as the pairing of an electron in one layer and a hole in another ( exciton condensation )
   ( see also Ref. \cite{fer} ). They also wrote down the wavefunctions for
   the ( smallest ) meron Eqn.\ref{meron}.
   Recently, Ye applied both Mutual Composite Fermion ( MCF) and Composite Boson (CB)
   Chern-Simons theories to study imbalanced $ \nu_{T} =1 $ bilayer \cite{prl,mcf} and
   trilayer quantum Hall systems \cite{tri}. He found that  CB approach is  more superior
    than the presently known  MCF approach in the study of interlayer coherent multi-layer QH systems.

   Because there is a gap in the bulk for generic 
   $ (m, m^{\prime}, n ) $ states, both the ground state wavefunctions  ( See Eqn.\ref{mn} ) and
   the quasi-hole wavefunctions ( see Eqn. \ref{qhmn} ) are good trial wavefunctions.
   However, in the interlayer coherent Quantum Hall $ (m, m, m) $ states, there is a neutral
   gapless mode even in the bulk. Earlier exact diagonalization studies \cite{he} showed
   that $ (111) $ wavefunction ( See Eqn.\ref{mn} ) remains a good one in
   the $ d \rightarrow 0 $ limit. However, as pointed out in \cite{rev},
   the $ (111) $ wavefunction may not be qualitatively good at finite $ d $,
   because $ (111) $ is a broken symmetry state of isotropic ferromagnet at $ d=0 $
   instead of a easy-plane ferromagnet at finite $ d $. Unfortunately, 
   a qualitatively good ground state wavefunction at finite $ d $ is still unknown \cite{mixed}. 
   In this paper, instead of finding such a good ground state wavefunction at finite $ d $,
   we focus on the quasi-hole wave functions
   ( see Eqn. \ref{qhmn} ) built on $ (111) $ state and measure its energy above the $ (111) $ state.
   So far their properties have not been investigated in detail.
   Their relation to meron wavefunctions in second quantization form is  not clear. For example,
   are they good trial wavefunctions ?  Are they new excitations different than
   merons ?
   The relation between $ (111) $ wavefunction and the EPQFM wavefunction in second quantization form
   is also not explicitly established.

  The gapless mode is a hallmark of the interlayer coherent Quantum Hall state.
  Its existence may dramatically alter the properties of the wavefunctions of the
  ground state, quasi-hole and quasi-particle.
  In this paper, we will study the relation between $ (111) $ wavefunction ( Eqn.\ref{111} ) and
  the EPQFM wavefunction ( Eqn.\ref{ex} ). We find that they are equivalent. We will also
  investigate the properties of meron wavefunction ( Eqn.\ref{meron} ) and the quasi-hole wave
  function ( see Eqn.\ref{qh} ). These two wavefunctions differ only by "normalization factors"
  ( the "normalization factor" is defined just above Eqn.\ref{sprime} ).
  Normalizations factors have been shown not to be important in single layer Quantum Hall ( SLQH )
  systems ( see section IIA).
  However, we find that they make a dramatic difference in BLQH. Although the smallest meron has
  a localized charge $ 1/2 $ and logarithmically divergent energy, 
  the charge of the quasi-hole excitation extends
  over the whole system and its energy also diverges linearly as the area of the system size.
  This indicates that the
  quasi-hole wavefunction is not a good trial wavefunction for any low energy excitations.
  We also find the energy of the possible wavefunction of a pair of merons ( Eqn.\ref{pair} ) written down in
  \cite{rev} increases quadratically instead of logarithmically as the separation of the pair increases.

    The rest of the paper is organized in the following. In Sec.~II, we will discuss the relations
   between the $ (111) $ wavefunction and the EPQFM wavefunction, the
   quasi-hole wavefunction and the meron wavefunction.
   In Sec.~III, we will study the charge distributions and energies of the meron and the quasihole, first by
   plasma analogy, then by Monte-Carlo simulations in both disk and spherical geometries. We will also
   discuss the properties of the trial wavefunction of a pair of merons.
   Finally, we state the conclusions in Sec.~IV.

\section{ Wavefunctions in Bilayer quantum Hall systems}

  We will first review some subtleties in the trial wavefunctions in single layer QH system,
  then point out these subtleties make important difference
  in the charge distributions and energies  of different trial wavefunctions in BLQH.

\subsection{ Subtleties in the trial wavefunctions in single layer QH system:}

   In symmetric gauge, the single-particle eigenfunctions of kinetic energy and angular momentum in the LLL
  are $ \phi_{m}(z) = \frac{1}{\sqrt{ 2 \pi 2^{m}  m ! }} z^{m} \exp( - |z|^{2}/4 ) $.

    At filling factor $ \nu=1/m $ where $ m=2s+1 $ is an odd integer, the Laughlin ground 
    state wavefunction is:
\begin{equation}
     \psi_{m}( z_{1}, \cdots, z_{N} ) = \prod^{N}_{j < k} ( z_{j}-z_{k} )^{m} e^{-\frac{1}{4} \sum_{i}
         |z_{i}|^{2} } =\sum_{ k_{1}, \cdots, k_{N} } C_{ k_{1}, \cdots, k_{N} }
                 z^{ k_{1} }_{1} \cdots z^{k_{N}}_{N} e^{-\frac{1}{4} \sum_{i} |z_{i}|^{2} } 
\label{lau}
\end{equation}
     where $ \sum^{N}_{i=1} k_{i}= \frac{ N (N-1) }{2} m  $ which is the total angular momentum of
     the state.
   
     If we insert one infinitesimally thin flux quantum at position $ z=0 $, then the single particle state
   is changed from $ \phi_{m}(z) $ to $ \phi_{m+1}(z) = \frac{z}{\sqrt{ 2 ( m+1) }} \phi_{m} ( z ) $.
    where the prefactor $ \frac{1}{\sqrt{ 2 ( m+1) }} $ will be called
    {\em normalization factor } in the following.
   The Quasi-hole wavefunction can be written as \cite{laug2} ( In the following, we will suppress the Gaussian factor ):
\begin{equation}
     \psi_{m}( z_{1},\cdots, z_{N} ) \rightarrow  \psi^{\prime}_{m,qh} (z) = \sum_{ k_{1}, \cdots, k_{N} }
       \frac{ C_{ k_{1}, \cdots, k_{N} } }{ \sqrt{ 2^{N} (k_{1}+1) \cdots ( k_{N}+1) }}
        z^{ k_{1}+1 }_{1} \cdots z^{k_{N}+1}_{N}
\label{sprime}
\end{equation}
     Removing all the normalization factors in front of the coefficients $ C_{ k_{1}, \cdots, k_{N} } $,
     the above wavefunction is simplified to
\begin{equation}
     \psi_{m}( z_{1},\cdots, z_{N} ) \rightarrow \psi_{m,qh} (z) = \sum_{ k_{1}, \cdots, k_{N} }
        C_{ k_{1}, \cdots, k_{N} }  z^{ k_{1}+1 }_{1} \cdots z^{k_{N}+1}_{N} = ( \prod_{i} z_{i} ) \psi_{m} 
\label{srm}
\end{equation}
    which is the well known Laughlin quasi-hole wavefunction. It carries localized fractional charge $ 1/m $.

  Note that for Integral Quantum Hall Effect (IQHE) $ m=1 $ when $ \psi_{1} $ is a single Slater determinant, it can be shown that
  all these normalization factors factor out of the determinant, therefore, removing
  these factors has no effect at all, namely, $ \psi^{\prime}_{m,qh} $ and $ \psi_{m,qh} $ are identical
  when $ m=1 $.
  However, for FQHE $ m > 1 $, $ \psi^{\prime}_{m,qh} $ and $ \psi_{m,qh} $
  are different.
   Obviously, both  $ \psi^{\prime}_{m,qh} $ and $ \psi_{m,qh} $ carry the same fractional charge
   $ \frac{1}{ 2m+1} $.
   In SLQH, it was shown by numerical studies in finite size that $ \psi_{m,qh} $
   and $ \psi^{\prime}_{m,qh} $ have similar energies \cite{norm}.
   However, $ \psi_{m,qh} $ reduces the density near the
   origin without ruining the good correlations in the ground state wavefunction $ \psi_{m} $, its energy 
   is expected to be slightly {\em lower} than that of $ \psi^{\prime}_{m,qh} $.

   $ \psi_{m,qh} $  can be best derived  from Jain's CF
   theory of SLQH system where the ground state wavefunction Eqn.\ref{lau} is rewritten as
   \cite{jain}.
\begin{equation}
     \psi_{m}( z_{1}, \cdots, z_{N} ) = \prod^{N}_{j < k} ( z_{j}-z_{k} )^{2s} \chi_{1} (z_{1}, \cdots, z_{N} )
\label{cfg}
\end{equation}
   where the $ 2s $ fold zeros $ ( z_{j}-z_{k} )^{2s} $ incorporates the strong correlations
   of electrons in the ground state and $ \chi_{1} $ is just the wavefunction of one filled Landau level.
    
   The wavefunction of a quasihole is:
\begin{eqnarray}
  \psi_{m,qh}  =   \prod^{N}_{j < k} ( z_{j}-z_{k} )^{2s} \chi_{1,hole},~~~~
   \chi_{1,hole}= ( \prod_{i} z_{i} ) \chi_{1}
\label{ex1}
\end{eqnarray}
   where $ \chi_{1,hole} $ is the wavefunction of a hole at $ \nu=1 $.

  If the QP and QH are far apart, the wavefunction for the QH at the origin is exactly  $ \psi_{m,qh} $
  instead of $ \psi^{\prime}_{m,qh} $, because inserting one flux quantum affects only $ \chi_{1} $ 
  without touching the correlation factor $ ( z_{j}-z_{k} )^{2s} $.

  Unfortunately, except for IQHE $ m=1 $, there are no {\em simple} second
  quantization forms for $ \psi_{m}, \psi_{m,qh}, \psi^{\prime}_{m,qh} $. We conclude that
  in SLQH system, the energies of trial wavefunctions are
  insensitive to whether or not we keep track of the normalization factors due to the existence of
  the gap in the bulk.
  In the following, we will apply the insights gained from SLQH systems to BLQH systems and arrive at
  very different conclusions.

\subsection{ $ 111 $ state in first and second quantization: }

    In first quantization, the ground state trial wavefunction of BLQH in $ d \rightarrow 0 $ limit
    is Halperin's $ (111) $ state \cite{bert}:
\begin{equation}
  \Psi_{111} (z,w)=  \prod^{N_{1}}_{i=1} \prod^{N_{2}}_{j=1} (z_{i}-w_{j} )
   \prod^{N_{1}}_{ i< j } ( z_{i}-z_{j} )   \prod^{N_{2}}_{ i< j } ( w_{i}-w_{j} )
\label{111}
\end{equation}
    where $ z $ and $ w $ are the coordinates in layer 1 and layer 2 respectively.

  In the EPQFM picture \cite{rev,fer}, the ground state EPQFM
  wavefunction was written in second quantization form:
\begin{equation}
  | G > =  \prod^{ M-1 }_{ m=0 } \frac{1}{ \sqrt{2} }
  ( C^{\dagger}_{m \uparrow} + C^{\dagger}_{m \downarrow} ) |0>
\label{ex}
\end{equation}
  where $ M=N $ is the angular momentum quantum number corresponding to the edge of the system.

   The wavefunction Eqn.\ref{ex} can also be interpreted
   as the pairing of an electron in one layer and a hole in another which leads to exciton condensation.
  In the following, we will show that the EPQFM wavefunction in the
  second quantization form of Eqn.\ref{ex} is equivalent to $ (111) $ wavefunction in Eqn.\ref{111}.

   We can expand Eqn.\ref{ex} into:
\begin{equation}
  | G > =  ( \frac{1}{ \sqrt{2} } )^{N} \sum^{N}_{ N_{1}=0 } \frac{1}{ N_{1} ! N_{2} ! }
  \sum_{ k_{1},\cdots, k_{N} }
  (-1)^{ P( k_{1},\cdots, k_{N}  ) } C^{\dagger}_{ k_{1}, \uparrow} \cdots C^{\dagger}_{ k_{ N_1}, \uparrow}
            C^{\dagger}_{k_{N_1+1}, \downarrow} \cdots  C^{\dagger}_{k_{N_1+ N_2},\downarrow} |0>
\end{equation}
   where $ N= N_1+ N_2 $ and $ P $ is the permutation of $ N $ variables.

   Projecting onto the state with $ N_{1} $ electrons in layer 1 and $ N_{2} $ electrons in layer 2
   ( this state will be denoted by $ (N_{1}, N_{2} ) $ in the following ) and then
   transforming into the first quantization form leads to:
\begin{eqnarray}
  | G; N_{1}, N_{2} > & = &  ( \frac{1}{ \sqrt{2} } )^{N}  \frac{1}{ N_{1} ! N_{2} ! }
  \sum_{ k_{1},\cdots, k_{N} } (-1)^{ P( k_{1},\cdots, k_{N}  ) } \frac{1}{ \sqrt{ N ! }}
  {\cal A}  [ \phi_{k_1}(z_{1}) \cdots \phi_{ k_{ N_1}}( z_{N_1} )  \nonumber  \\
   &    &  \phi_{k_{ N_1+1}} (w_{1}) \cdots \phi_{ k_N}( w_{N_2} )
   (1, \uparrow ) \cdots (N_1, \uparrow ) ( N_1+1, \downarrow ) \cdots ( N, \downarrow) ]
\end{eqnarray}
   where $ {\cal A} $ stands for anti-symmetrization.

   Moving the summation over the orbital states into $ {\cal A} $, the sum
   is essentially the $ (111) $ state:
\begin{eqnarray}
   & &\sum_{ k_{1},\cdots, k_{N} } (-1)^{ P( k_{1},\cdots, k_{N}  ) }
   \phi_{k_1}(z_{1}) \cdots \phi_{ k_{ N_1}} ( z_{N_1} )
   \phi_{k_{ N_1+1}} (w_{1}) \cdots \phi_{ k_N}( w_{N_2} )   \nonumber  \\
   & & ~~~~~= \prod^{N_{1}}_{i=1} \prod^{N_{2}}_{j=1} (z_{i}-w_{j} )
   \prod^{N_{1}}_{ i< j } ( z_{i}-z_{j} )   \prod^{N_{2}}_{ i< j } ( w_{i}-w_{j} ) =\psi_{111}(z,w)
\end{eqnarray}
   Finally
\begin{equation}
  | G; N_{1}, N_{2} > =  ( \frac{1}{ \sqrt{2} } )^{N}  \frac{1}{ N_{1} ! N_{2} ! } \frac{1}{ \sqrt{N !}}
  {\cal A}  [ \psi_{111}( z,w)
   (1, \uparrow ) \cdots (N_1, \uparrow ) ( N_1+1, \downarrow ) \cdots ( N, \downarrow) ]
\end{equation}

   This concludes that the orbital part of the projection of the EPQFM wavefunction Eqn.\ref{ex}
   onto $ ( N_{1}, N_{2} ) $ sector is exactly the same as $ (111) $ state. In the following, we will
   study the excitations of the $ (111) $ state.

\subsection{ " Quasi-hole " versus Meron wavefunction:}

   In first quantization, the most obvious trial wavefunctions for the "Quasi-hole"
   at the origin in layer 1 and layer 2 are \cite{rev}:
\begin{equation}
  \Psi_{qh,1}(z,w) =  ( \prod^{N_{1}}_{i=1} z_{i} ) \Psi_{111},~~~~~~~~
  \Psi_{qh,2}(z,w) =  ( \prod^{N_{2}}_{i=1} w_{i} ) \Psi_{111}
\label{qh}
\end{equation}

   We can get the above trial wavefunctions by naively generalizing
  the composite fermion quasi-hole wavefunction Eqn.\ref{ex1} in SLQH to
  the corresponding quasi-hole wavefunction in BLQH:
\begin{equation}
  \Psi_{qh,1}(z,w) = \prod^{N_{1}}_{i=1} \prod^{N_{2}}_{j=1} (z_{i}-w_{j} ) 
   \prod^{N_{2}}_{ i< j } ( w_{i}-w_{j} ) \times ( one~ hole~ in~ layer~ 1 ) 
\label{ex2}
\end{equation}

  In second quantization, the meron wavefunction in EPQFM picture  \cite{rev},
  created by inserting one flux quantum in layer 2 at the origin, is
\begin{equation}
  |Meron,2 > =  \prod^{ M-1 }_{ m=0 } \frac{1}{ \sqrt{2} }
  ( C^{\dagger}_{m \uparrow} + C^{\dagger}_{m+1 \downarrow} ) |0>
\label{meron}
\end{equation}
  where $ 0, \downarrow $ state is not occupied.
 
  An important question to ask is whether the second quantization form of the meron
  wavefunction is equivalent to the  "quasi-hole" wavefunction in the first quantization form in Eqn.\ref{qh}. 
 
  In the following, we will show that they are different.

  We can expand Eqn.\ref{meron} into:
\begin{equation}
  | Meron,2 > =  ( \frac{1}{ \sqrt{2} } )^{N} \sum^{N}_{ N_{1}=0 } \frac{1}{ N_{1} ! N_{2} ! }
  \sum_{ k_{1},\cdots, k_{N} }
  (-1)^{ P( k_{1},\cdots, k_{N}  ) } C^{\dagger}_{ k_{1}, \uparrow} \cdots C^{\dagger}_{ k_{ N_1}, \uparrow}
            C^{\dagger}_{k_{N_1+1} +1, \downarrow} \cdots  C^{\dagger}_{k_{N_1+ N_2}+1,\downarrow} |0>
\end{equation}

   Projecting onto the state $ ( N_{1}, N_{2} ) $ and then
   transforming the above equation into the first quantization form lead to:
\begin{eqnarray}
  | meron, 2; N_{1}, N_{2} > & = &  ( \frac{1}{ \sqrt{2} } )^{N}  \frac{1}{ N_{1} ! N_{2} ! }
  \sum_{ k_{1},\cdots, k_{N} } (-1)^{ P( k_{1},\cdots, k_{N}  ) } \frac{1}{ \sqrt{ N ! }}
  {\cal A}  [ \phi_{k_1}(z_{1}) \cdots \phi_{ k_{ N_1}}( z_{N_1} )  \nonumber  \\
   &    &  \phi_{k_{ N_1+1} +1 } (w_{1}) \cdots \phi_{ k_N +1 }( w_{N_2} )
   (1, \uparrow ) \cdots (N_1, \uparrow ) ( N_1+1, \downarrow ) \cdots ( N, \downarrow) ]
\end{eqnarray}

    Using $ \phi_{m+1}(w) = \frac{w}{\sqrt{ 2 ( m+1) }} \phi_{m} ( w ) $ and
    moving the summation over the orbital states into $ {\cal A} $, we find
    the orbital part becomes
\begin{eqnarray}
  \Psi_{meron,2} ( z, w ) & = & \sum_{ k_{1},\cdots, k_{N} } (-1)^{ P( k_{1},\cdots, k_{N}  ) }
   \phi_{k_1}(z_{1}) \cdots \phi_{ k_{ N_1}} ( z_{N_1} )   \nonumber   \\
  & \times &  \frac{ w_{1}}{ \sqrt{ 2 ( k_{ N_1+1} +1 ) } } \phi_{k_{ N_1+1}} (w_{1}) \cdots
   \frac{ w_{N_2}}{ \sqrt{ 2 ( k_{ N} +1 ) } } \phi_{ k_N}( w_{N_2} )
\label{coe} 
\end{eqnarray}

  Naively, one may expect that all the normalization factors can be factored out just like in
  $ m=1 $ IQHE in SLQH. However, detailed calculations show that this is not the case.
  In fact, these normalization factors play an important role.

   If we neglect the normalization factors just like what
   we did in SLQH in Eqn.\ref{srm},
   the above sum leads simply to the quasi-hole wavefunction in layer 2 listed in Eqn.\ref{qh}.
\begin{equation}
    \Psi_{qh,2} ( z, w ) = ( \prod^{N_{2}}_{i=1}  w_{i} ) \Psi_{111} ( z, w )
\end{equation}

   If keeping all the normalization factors, then we note that $ \phi_{m}(z) $ peaks strongly around its maximum at
   $ |z|^{2}= 2 m $. 
   If $ | w_{i} | \gg 1 $, we can replace $ \frac{w_{i}}{ \sqrt{ 2 ( k_{N_1+i} +1 ) }} $
   in Eqn.\ref{coe} by $ \frac{w_{i}}{ | w_{i} | } $.

   Therefore, far away from the origin, we have: 
\begin{equation}
   \Psi_{meron,2} ( z, w \rightarrow \infty ) =
   ( \prod^{N_{2}}_{i=1}  \frac{w_{i}}{ | w_{i} | }  ) \Psi_{111} ( z, w )
\label{far}
\end{equation}

    Let us look at the following complete wavefunction which includes both spin part and orbital part
\begin{equation}
       \Psi = \prod_{i} \frac{1}{ \sqrt{2} } \left ( \begin{array}{c}
        1   \\
        Z_{i}/|Z_{i}|  \\
   \end{array}   \right ) \Psi_{1}( Z )
\label{complete}
\end{equation}
     where $ Z=( z, w ) $.

     Expanding and then projecting onto $ ( N_{1}, N_{2} ) $ sector and extracting the orbital part,
     we can see the orbital part is exactly the same as Eqn.\ref{far}.

    Setting $ Z_{i} = |Z_{i}| e^{i \theta_{i} } $, the spin part of
    the wavefunction can be written as \cite{rev}:
\begin{equation}
       \chi ( \theta ) =  \frac{1}{ \sqrt{2} } \left ( \begin{array}{c}
        1   \\
        e^{i \theta}  \\
   \end{array}   \right ) 
\label{spin}
\end{equation}
    which is clearly a vortex.

   It was shown that the meron wavefunction Eqn.\ref{meron} or \ref{coe} has logarithmically divergent
   energy with the system size $ E_{m} \sim \ln M $ 
   where $ M=N $ is the angular momentum quantum number corresponding to the edge of the system \cite{rev}.

  This shows that the orbital part of the projection of the second quantized form of meron wavefunction
  Eqn.\ref{meron} onto $ ( N_{1}, N_{2} ) $ sector is {\em not } the same as the quasi-hole wavefunction 
  Eqn.\ref{qh}. The properties of the "quasi-hole "
  wavefunction Eqn.\ref{qh} have not been seriously studied so far. In the following, we will
  study the charge distributions and energy of this wavefunction and find that they are quite different
  from those of meron wavefunctions. 

\section{ Investigation of the energy of quasi-hole wavefunction }

    In this section, from plasma analogy,
    we will study the charge distributions and energies of quasi-hole wavefunctions
    qualitatively and gain some physical insights. Then we will study them quantitatively
    by Monte-Carlo simulations in both disk and spherical geometries.
    We will also compare with the meron wavefunctions. 

\subsection{ Plasma analogy of the quasi-hole wavefunction}

    We will first review Halperin's $ (m, m^{\prime}, n) $ state \cite{bert}
    and then study the very interesting case $ (111) $ state which is a interlayer coherent state
    in this series. The $ (m, m^{\prime}, n) $ wave function is given by:
\begin{equation}
  \Psi_{ m, m^{\prime}, n } (z,w)=  \prod^{N_{1}}_{i=1} \prod^{N_{2}}_{j=1} (z_{i}-w_{j} )^{n}
   \prod^{N_{1}}_{ i< j } ( z_{i}-z_{j} )^{m}  \prod^{N_{2}}_{ i< j } ( w_{i}-w_{j} )^{m^{\prime}}
\label{mn}
\end{equation}

    The quasi-hole wavefunctions were written down in \cite{rev}:
\begin{equation}
  \Psi_{qh,1}(z,w) =  ( \prod^{N_{1}}_{i=1} z_{i} ) \Psi_{m, m^{\prime}, n},~~~~~~~~
  \Psi_{qh,2}(z,w) =  ( \prod^{N_{2}}_{i=1} w_{i} ) \Psi_{m, m^{\prime}, n}
\label{qhmn}
\end{equation}

  These trial wave functions clearly reduce the density near the origin without ruining the good correlations
  in the ground state $  (m, m^{\prime}, n) $.
  The fractional charges of these quasi-hole states were studied in
  \cite{rev} from two component plasma analogy. From the perfect screening conditions, the fractional
   charges of $ \Psi_{qh,1}(z,w) $ in layer 1 ( $ e^{*}_{1} $ ) and in layer 2 ( $ e^{*}_{2} $ )
   are found to be:
\begin{equation}
  e^{*}_{1}=\frac{m^{\prime}}{ m m^{\prime} - n^{2} },~~~e^{*}_{2}=\frac{-n}{ m m^{\prime} - n^{2} },~~~
  e^{*}_{t}=\frac{m^{\prime}-n}{ m m^{\prime} - n^{2} },
  ~~~e^{*}_{r}=\frac{m^{\prime}+n}{ m m^{\prime} - n^{2} }
\end{equation}
   where $ e^{*}_{t}, e^{*}_{r} $ are the total and relative charges.
   
   It is easy to see $ e^{*}_{1} $ and $ e^{*}_{2} $ always carry opposite charges as expected because
   of the strong interlayer correlations built in the $ ( m, m^{\prime}, n ) $ ground state wavefunction.
   It is also easy to see $ e^{*}_{t} = \nu_{1} $ where  $ \nu_{1} $ is the filling factor in layer 1.
   The fractional charges for $ \Psi_{qh,2}(z,w) $ can be obtained from the above results by
   $ m \rightarrow m^{\prime},  \nu_{1} \rightarrow \nu_{2} $ where $ \nu_{2} $ is the filling factor
   in layer 2.

    For the $ (111) $ ground state, it is easy to see both $ e^{*}_{1} $ and $ e^{*}_{2} $ diverge,
    while $  e^{*}_{t} = \nu_{1} $ remains finite to be equal to the filling factor in layer 1.
    In the balanced case $ \nu_{1} = \nu_{2} =1/2 $,  $  e^{*}_{t}=1/2 $, namely, the total charge
    is 1/2 \cite{mis}. However, $ e^{*}_{r} $ diverges. It is well known that there is a capacitive
    energy term $ E_{c} \sim \int d^{2} \vec{r} ( \rho_{2} - \rho_{1} )^{2} $ for BLQH with finite distance $ d $.
    This term strongly suppress the relative density fluctuations between the two layers. Therefore,
    the divergence of $ e^{*}_{r} $ indicates that the energy of quasi-hole may diverge much faster
    than that of meron wavefunction which diverges only logarithmically \cite{edge}.

    In the following, we will study the charge distributions and energy quantitatively 
    by Monte-Carlo simulations in both disk and spherical geometry.
\subsection{ Monte-Carlo simulations}

   The charge density operator is defined as $ \rho(z) = \sum^{N}_{i=1} \delta( z-z_{i} ) $.
   In SLQH system, its average value at $ z $ is 
\begin{equation}
  \rho(z)= N \int d^{2} z_{2} \cdots d^{2} z_{N} | \psi( z, z_{2}, \cdots, z_{N} ) |^{2}/Z,
\end{equation}
   where $  Z =\int d^{2} z | \psi( z ) |^{2} $
   and the  normalization is $ \int d^{2} z \rho(z) = N $.

   For BLQH system, the charge densities at the two layers $ \rho_{1}(z), \rho_{2}(w) $ are:
\begin{eqnarray}
  \rho_{1}(z) & = &  N_1  \int d^{2} z_{2} \cdots d^{2} z_{N_1}
              \int d^{2} w | \psi( z, w ) |^{2}/Z
                          \nonumber  \\
  \rho_{2}(w) & = &  N_2  \int d^{2} z
              \int d^{2} w_{2} \cdots d^{2} w_{N_2} | \psi( z, w ) |^{2}/Z
\end{eqnarray}
    with $ Z = \int d^{2} z d^{2} w | \psi ( z, w ) |^{2} $ and
    the normalizations $  \int d^{2} z \rho_{1}(z) = N_{1}, \int d^{2} w \rho_{2}(w) = N_{2} $.

   In the following, we  focus on the balanced case $ N_{1}= N_{2}= N/2 $.

   For the ground state $ (111) $ wavefunction, we find:
\begin{equation}
  \rho_{10}(z)= \rho_{20}(z) = \rho_{0}(z) = \frac{1}{2} \sum^{M-1}_{0} | \phi_{m} ( z) |^{2}
\end{equation}

    While, for the meron wavefunction, we find:
\begin{equation}
  \rho_{1m}(z)= \frac{1}{2} \sum^{M-1}_{0} | \phi_{m} ( z) |^{2} = \rho_{0}(z),~~~~~~~
  \rho_{2m}(z)= \frac{1}{2} \sum^{M}_{1} | \phi_{m} ( z) |^{2}
\end{equation}
 
    The deviations from the ground state in the two layers are
\begin{equation}
  \delta \rho_{1m}(z) = \rho_{1m}(z) - \rho_{0}(z) = 0,~~~~~~~
  \delta \rho_{2m}(z) = \rho_{2m}(z) - \rho_{0}(z) = - \frac{1}{2}( |\phi_{0}(z) |^{2} - |\phi_{M}(z) |^{2})
\end{equation}
    The charges in top layer remains untouched,
    while 1/2 charge is moved from the origin to the boundary ( Fig.1 ),
    namely, there is a hole with charge 1/2 at the origin and extra charge -1/2 at the
    boundary.
     
\vspace{0.5cm}

\epsfig{file=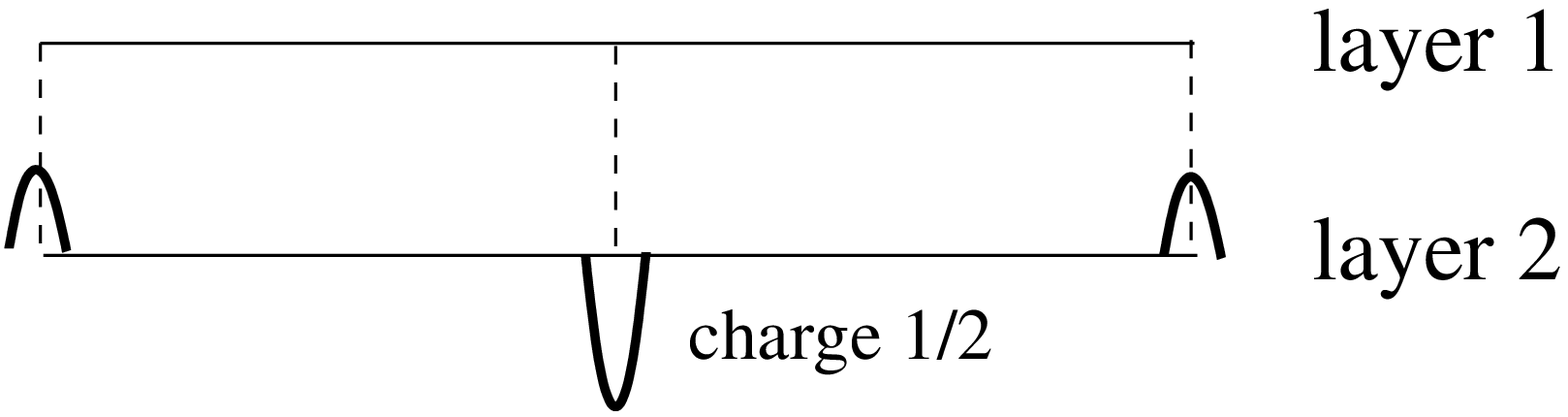,width=2.5in,height=1.0in,angle=0}

\vspace{0.25cm}

{\footnotesize {\bf Fig 1:} The charge densities of the smallest meron wavefunction }

\vspace{0.25cm}

    The fact that the charges in top layer remains unmodified when one flux quantum
    is inserted in the bottom layer is because
    the meron wavefunction ignores the strong interlayer correlation. We expect
    the neutral gapless mode will strongly modify the charge distributions of a meron.


     We can compare the above meron wavefunction to the wavefunction of the ordinary charge 1 excitation:
\begin{equation}
  \Psi_{ch1}(z,w) =  ( \prod^{N_{1}}_{i=1} z_{i} \prod^{N_{2}}_{i=1} w_{i} ) \Psi_{111}
\label{charge1}
\end{equation}
    Obviously, it has finite energy. Its charge distribution is depicted in Fig.2.
\vspace{0.5cm}

\epsfig{file=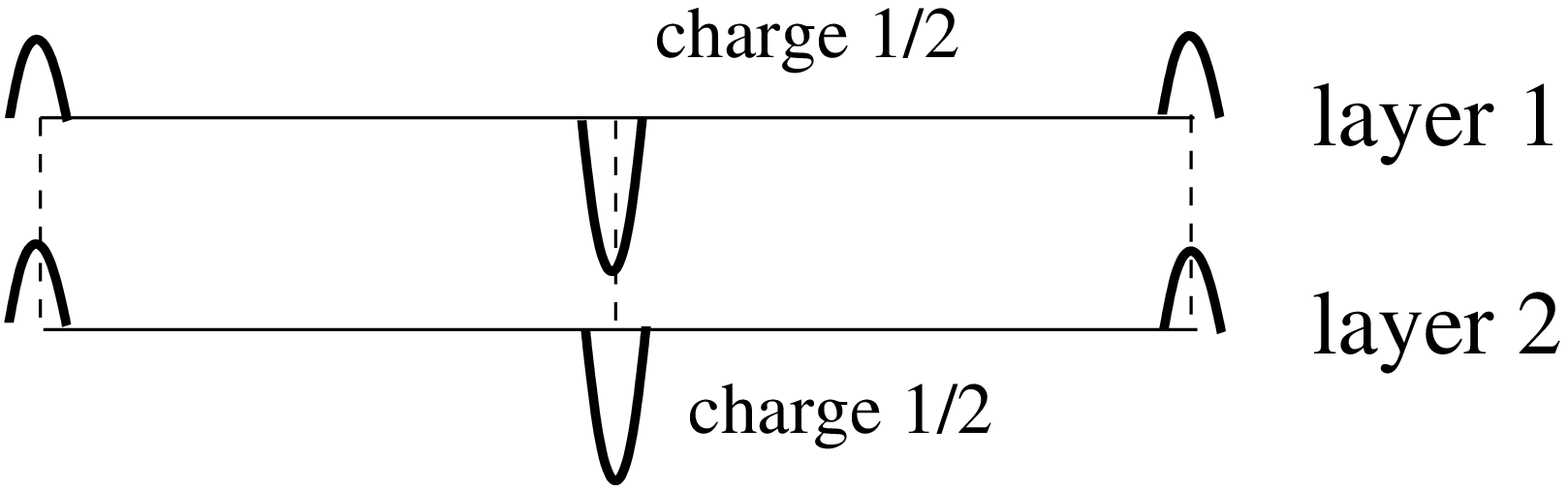,width=2.5in,height=1.0in,angle=0}

\vspace{0.25cm}

{\footnotesize {\bf Fig 2:} The charge densities of the charge 1 excitation }

\vspace{0.25cm}

   The density profile for the quasi-hole in layer 2 is:
\begin{eqnarray}
  \delta \rho_{1qh}(z) & =  &   
   \frac{N}{2}  \frac{ \sum_{ m_{1},\cdots, m_{N} }  |\phi_{m_{1}}(z) |^{2} ( m_{ N_1+1} +1 ) \cdots
   ( m_{ N} +1 ) }{ \sum_{ m_{1},\cdots, m_{N} }( m_{ N_1+1} +1 ) \cdots ( m_{ N} +1 ) }
    -\frac{1}{2} \sum^{M-1}_{0} | \phi_{m} ( z) |^{2}   \nonumber  \\ 
 \delta \rho_{2qh}(z) & =  &   
   \frac{N}{2}  \frac{ \sum_{ m_{1},\cdots, m_{N} }  |\phi_{m_{ N_1 +1 } + 1}(z) |^{2} ( m_{ N_1+1} +1 ) \cdots
   ( m_{ N} +1 ) }{ \sum_{ m_{1},\cdots, m_{N} }( m_{ N_1+1} +1 ) \cdots ( m_{ N} +1 ) }
    -\frac{1}{2} \sum^{M-1}_{0} | \phi_{m} ( z) |^{2}  
\end{eqnarray}
   The above expressions can be evaluated analytically only for small number of particles.
   Fortunately, Monte-Carlo simulations can produce various quantities for large number of particles.
   In the following, we will first discuss the results in disk geometry, then in spherical geometry.

\subsubsection{ Results in disk geometry }

\vspace{0.5cm}

\epsfig{file=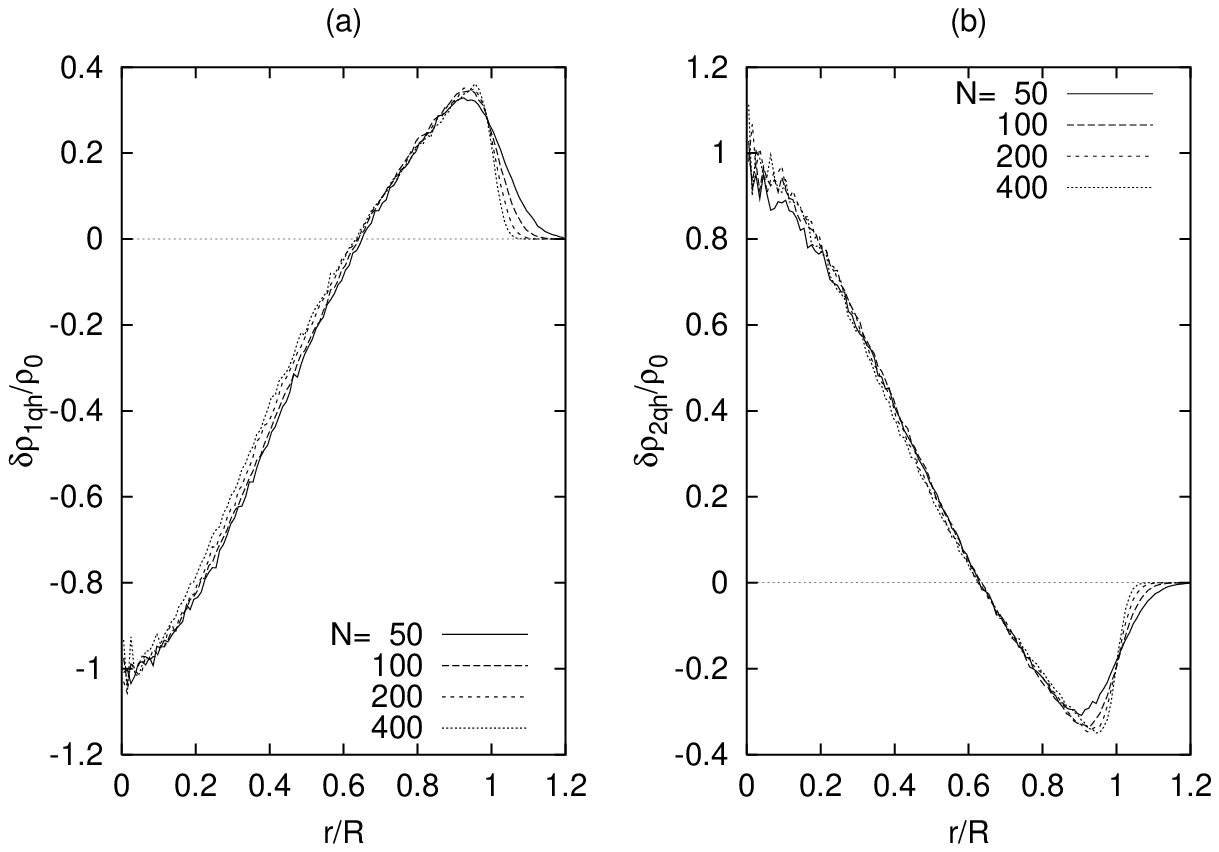,width=3.5in,height=2.0in,angle=0}

\vspace{0.5cm}

{\footnotesize {\bf Fig 3:} The charge densities of the quasi-hole wavefunction in layer 1 and 2 }

\vspace{0.25cm}

   As shown in Fig.3, inserting one flux quantum at bottom layer pushes electrons towards the boundary
   to create charge deficits near the origin at the bottom layer. At the same time, electrons are pulled
   from the boundary to create charge surplus near the origin on the top layer. However, 
   the excitations are not localized but extends over the whole system.
   Both density plots seem to collapse roughly on a single
   curve (except around the boundary) when the distance is scaled by the system size $ R=\sqrt{2N/\nu}$,
   namely, $ \delta \rho_{1qh} (r) = f_{1} (r/R), \delta \rho_{2qh} (r) = f_{2} (r/R) $.
   We can see at $ r/R \sim 0.64 $ both density deviations change sign. This value is close to the radius
   $ r/R = 1/\sqrt{2} \sim 0.7 $ where the area inside is equal to the area outside in a disk geometry.
   We may tentatively define this radius as
   the size over which the quasi-particle charge extends.

   We can evaluate the total "quasi-hole" charge $ \rho_{qht}= \delta \rho_{1qh} + \delta \rho_{2qh} $
   and charge density difference profile
   $ \delta \rho_{qh} = \delta \rho_{1qh}- \delta \rho_{2qh} $ between the two layers.
 They are shown in Fig. 4a  and Fig. 4b respectively.
 Both  extend over the whole system. Again $ \delta \rho_{qh} = f_{1} (r/R)- f_{2} (r/R) =  f(r/R) $
 changes sign at $ r/R \sim 0.64 $. The fact that all $ \rho_{qht} $ curves approach
 the horizontal axis as $ N $ increases is expected, because the total area enclosed by these curves should be
 the total charge $ -1/2 $ which spreads over the whole system, therefore the total charge density
 approaches zero as $ N $ increases.

\vspace{0.5cm}

\epsfig{file=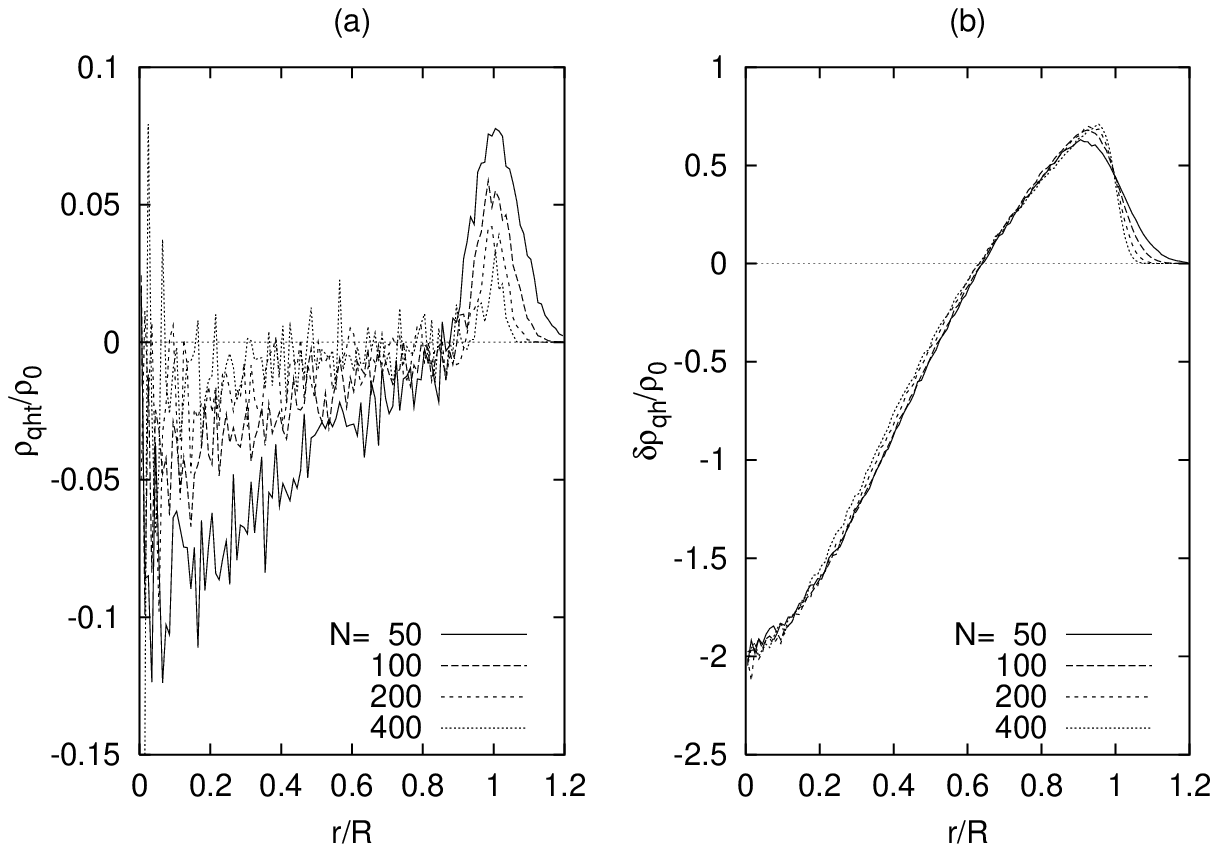,width=3.5in,height=2.0in,angle=0}

\vspace{0.5cm}

{\footnotesize {\bf Fig 4:} The total and relative charge densities of the quasi-hole wavefunction }

\vspace{0.25cm}

   The integrated total and relative charges inside the radius $r$, i.e.
   $ q_i(r) = \int_{r'<r} d^2 r' [\rho_i(r') - \rho_0(r')], i=1,2 $ are shown in Fig.5a and 5b.

\vspace{0.5cm}

\epsfig{file=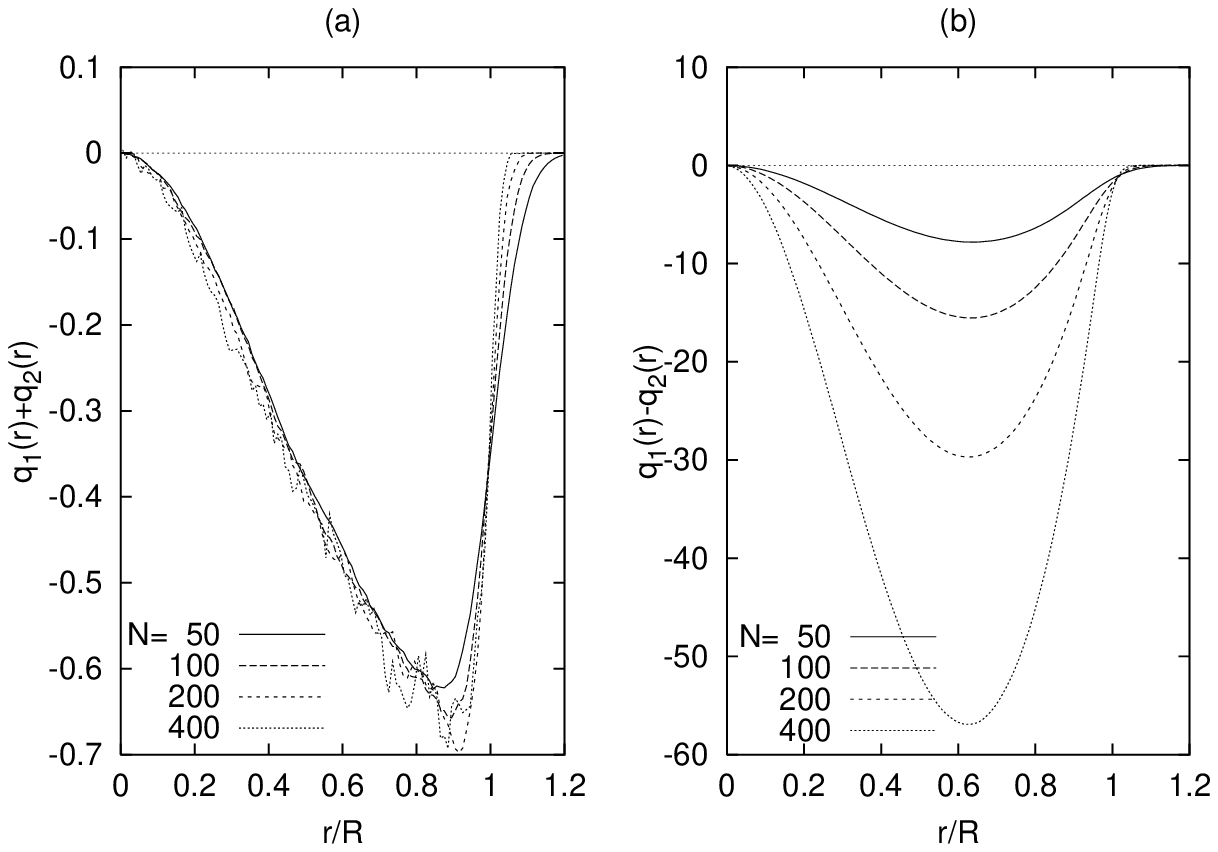,width=3.5in,height=2.0in,angle=0}

\vspace{0.5cm}

{\footnotesize {\bf Fig 5:} The integrated total and relative charges inside radius $ r $.}

\vspace{0.25cm}

     As shown in Fig.5, the total integrated charges at $ r/R \sim 0.64 $ are indeed -1/2, which is consistent with
   the result achieved in the plasma analogy. However they do not show a plateau
   because there is no plateau in the relative density ( Fig. 3) either. The integrated relative charges
   simply diverge linearly as the system size grows. This is expected, because $ Q(r)=
   q_{1}(r) - q_{2}(r) = 2 \pi \int^{r}_{0}  f ( r^{\prime}/R ) r^{\prime} d r^{\prime}   
   = 2 \pi R^{2} \int^{r/R}_{0} f(x) x dx = R^{2} F_{Q}(r/R) $. Setting $ r_{0}/R \sim 0.64 $,
   we get $ Q(r_{0}) \sim R^{2} \sim N $. The plasma analogy implies the total charge difference
   $ Q $ diverges, the Monte-Carlo simulations explicitly show that $ Q $ diverges linearly as $ N $.
   We can also estimate the energy of this quasi-hole by first estimating the capacitive
   energy $ E_{c} = \frac{ Q^{2} }{ 2 C }  $ where 
   $ C $ is the capacitance which is inverse proportional to the distance $ d $ between the two layers
   and proportional to the area the charge spreads over. From the above analysis $ Q \sim N, C \sim N $,
   then $ E_{c} \sim N $. We expect the energy of the quasi-hole should be at the same order as 
   $ E_{c} $, therefore $ E_{qh} \sim N $, namely, the quasi-hole energy diverges linearly as
   the system size grows. This linear divergence is much faster than the logarithmic divergence
   of the meron energy.

   We also show the quantitative capacitive energy $ E_{c}=  \int d^{2} \vec{r} ( \rho_{2} - \rho_{1} )^{2} $
   in Fig.6
\vspace{0.5cm}

\epsfig{file=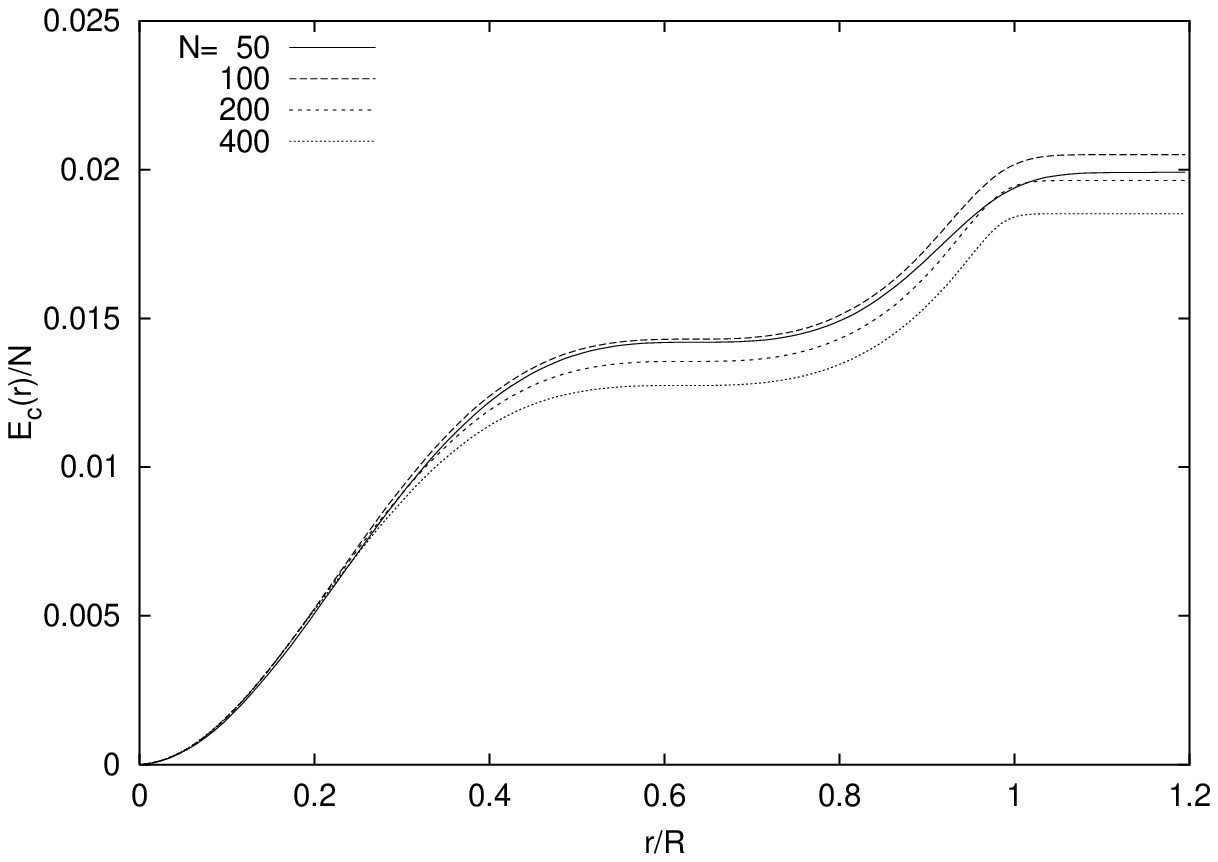,width=3.5in,height=2.0in,angle=0}

\vspace{0.5cm}

{\footnotesize {\bf Fig 6:} The capacitive energy of BLQH }

\vspace{0.25cm}

   In Fig, 6 is shown that $ E_{c} ( r_{0}/R ) \sim 0.015 N $, namely, increase linearly as $ N $.
   This is expected, because $ E_{c} ( r ) =
   2 \pi \int^{r}_{0} f^{2} ( r^{\prime}/R ) r^{\prime} d r^{\prime}  
   = 2 \pi R^{2} \int^{r/R}_{0} f^{2}(x) x dx = R^{2} F_{c}(r/R) $, Setting $ r_{0}/R \sim 0.64 $,
   we get $ E_{c}(r_{0}) \sim R^{2} \sim N $ which is consistent with the above estimate of $ E_{c} $.
   The plasma analogy implies the quasi-hole energy $ E_{qh} $ diverges, the Monte-Carlo simulations explicitly
   show that $ E_{qh}  $ diverges linearly as $ N $.

   Note that Eqn.\ref{charge1} is essentially the wavefunction of a charge 1 excitation in single layer
   spin polarized $ \nu=1 $ Quantum Hall state.
   It can also be considered as a smallest
   skyrmion. When we split the skyrmion,
   namely, replacing the prefactor
   $ ( \prod^{N_{1}}_{i=1} z_{i}  \prod^{N_{2}}_{i=1} w_{i} ) $ in Eqn.\ref{charge1} by
   $  \prod^{N_{1}}_{i=1} (z_{i} - z_{0})  \prod^{N_{2}}_{i=1} ( w_{i}-w_{0} ) $ and then
   gradually increasing $ | z_{0} -w_{0} | $,
   its energy $ E_{ch1} \sim | z_{0} -w_{0} |^{2} $ increases quadratically.
   Since, in the $ | z_{0} -w_{0} | \rightarrow \infty $ limit,
   Eqn.\ref{charge1} will reduce to the two charge 1/2  quasi-hole wave functions in Eqn.\ref{qh}.
   
    An approximate wavefunction for a pair of meron which includes both spin and
    orbital part was proposed in Eqn.208 of \cite{rev}:
\begin{equation}
       \Psi_{pair} = \prod_{i} \frac{1}{ \sqrt{2} } \left ( \begin{array}{c}
        z_{i}- z_{0}   \\
        w_{i} -w_{0}  \\
   \end{array}   \right ) \Psi_{1}( Z )
\label{pair}
\end{equation}
     where $ Z=( z, w ) $.

   As discussed in \cite{rev}, the spin is up at $ w_{0} $ and down at $ z_{0} $. By symmetry,
   it seems that there may be $1/2 $ charge located at $ w_{0} $ and $  z_{0} $, so it may be a good
   candidate for the wavefunction of a pair of merons.
    By a unitary transformation ( Eqn. (110) in Ref.\cite{rev} ), Eqn.\ref{pair} can be mapped to
    a skymion wavefunction with a finite radius. This shows that at zero distance,
    the candidate wavefunction for a pair of merons  Eqn.\ref{pair} has the same energy
    as that of a skymion and one  meron can be viewed as half of a skymion.
    However, at finite distance, the Hamiltonian has only $ U(1) $ symmetry,
    the unitary transformation does not respect this $ U(1) $ symmetry anymore.
    Eqn.\ref{pair} has much lower energy than that of a skymion, since it keeps the spins in the $ XY $ plane.

    Now let's look at the energy of Eqn.\ref{pair}.
   It is easy to see that the projection of this wavefunction to a given sector $ (N_{1}, N_{2} ) $
   is $ \prod^{N_{1}}_{i=1} (z_{i} - z_{0})  \prod^{N_{2}}_{i=1} ( w_{i}-w_{0} ) \Psi_{111} $, so
   its energy $ E_{pair} \sim | z_{0} -w_{0} |^{2} $ instead of logarithmically as naively expected,
   because the charges are extended between $ z_{0} $ and $ w_{0} $. A good wavefunction
   for a pair of meron whose energy increases logarithmically as the distance between the two merons
   is still unknown.

   We can also calculate directly the energy of the (111) state $E_0$ and the quasihole state $E_{qh}$.
   The intra- and inter-layer Coulomb interactions are given by:
\begin{equation}
  V_{11}= V_{22} = 1/r,~~~~~ V_{12}=V_{21}= 1/\sqrt{r^2+d^2}
\end{equation}
where $d$ is the distance between the layers.

  Unfortunately, the boundary contributions from the positive background charges
  in the disk geometry are not easily taken care of. In order to  avoid this difficulty, we 
  perform the simulations in spherical geometry.

Through the use of the single-particle eigenfunctions 
 $\phi_{q,m} \propto u^{q+m} v^{q-m} e^{i q \phi}$
 in the lowest Landau level, 
we can construct Halperin's $ (111) $ wavefunction in spherical geometry :
\begin{equation}
  \Psi_{111} (z,w)=  \prod^{N_{1}}_{i=1} \prod^{N_{2}}_{j=1}
  [ ( u_{i} \tilde{v}_{j}-v_{i} \tilde{u}_{j} ) e^{i(\phi_i +\tilde{\phi}_j)}]
   \prod^{N_{1}}_{ i< j } 
	[ ( u_{i} v_{j}-v_{i} u_{j} )  e^{i(\phi_i +\phi_j)} ]
	\prod^{N_{2}}_{ i< j }
	[ ( \tilde{u}_{i} \tilde{v}_{j}- \tilde{v}_{j} \tilde{u}_{i} )
	e^{i(\tilde{\phi}_i +\tilde{\phi}_j)} ]
\label{111sp}
\end{equation}
    where $ z= ( u, v ) $ with $ u= \cos \frac{\theta}{2} e^{-i \phi/2}, v= \sin \frac{\theta}{2} e^{i \phi/2}  $
    are the spinor coordinates in
    sphere 1 and $ w=( \tilde{u}, \tilde{v} ) $
     with $ \tilde{u} =  \cos \frac{ \tilde{\theta} }{2} e^{-i \tilde{\phi}/2}, \tilde{v} =
     \sin \frac{ \tilde{\theta} }{2} e^{i \tilde{\phi}/2}  $ are the spinor coordinates
    in sphere 2 \cite{int}.

   The quasi-hole wavefunction at north pole in sphere 1 is
\begin{equation}
  \Psi_{qh,1n}(z,w) =  ( \prod^{N_{1}}_{i=1} v_{i} e^{i\phi_i/2} ) \Psi_{111}
\label{qhsph}
\end{equation}
    which corresponds to increase the monopole strength $ Q $ in the center of the sphere 1 by $ 1/2 $,
    namely, $ Q \rightarrow Q + 1/2 $. 

  We evaluate the energy difference $ E_{h} - E_{0} $ in spherical geometry where the background
  positive charge distributions cancel. The results at several finite separations are shown in Fig. 7.

\vspace{0.5cm}

\epsfig{file=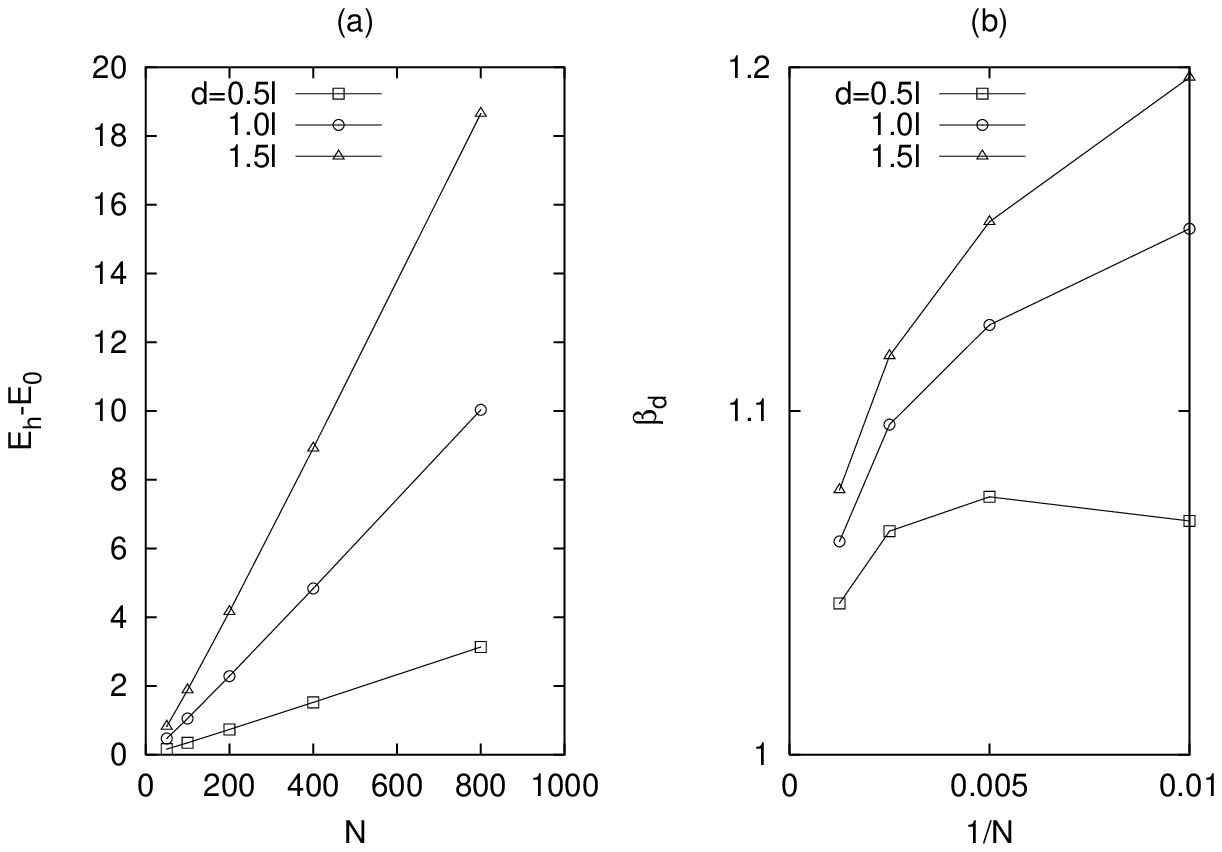,width=3.5in,height=2.5in,angle=0}

\vspace{0.5cm}

{\footnotesize {\bf Fig 7:} The energies of quasi-hole wavefunction at different separations $ d > 0 $
   of the two layers in spherical geometry }

\vspace{0.25cm}

   As shown in Fig. 7a, $ E_{h}- E_{0} = C(d) N $ with $ C(d) \sim d \rightarrow 0 $ as $ d \rightarrow 0 $.
   We also simulate the best fit for the power law behavior $ E_{h}- E_{0} = C(d) N^{\beta_{d} } $
   at a given $ d $ as the system size $ N $ increases. As shown in Fig. 7b, the power
   $ \beta_{d}( N ) $ are slightly larger than one in finite size, but all approach to one in the
   $ N \rightarrow \infty $ thermodynamic limit.

   We also calculate $ E_{h}- E_{0} $ at $ d=0 $. The result is shown in Fig.8.
   
\vspace{0.5cm}

\epsfig{file=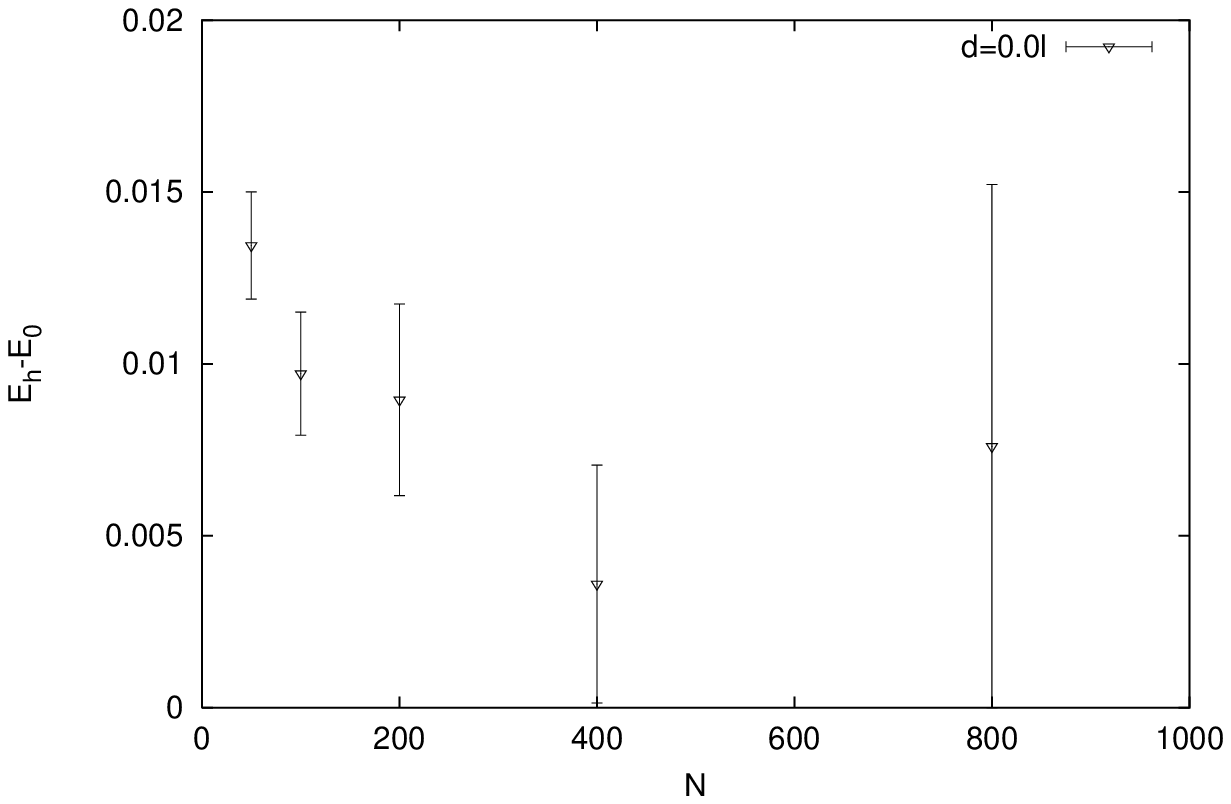,width=3.5in,height=2in,angle=0}

\vspace{0.5cm}

{\footnotesize {\bf Fig 8:} The energy of the quasi-hole wavefunction at $ d = 0 $ in spherical geometry }

\vspace{0.25cm}

    In sharp contrast to all the results at $ d > 0 $ shown in Fig. 7,
   the energy at $ d=0 $ remains finite in the thermodynamic limit.
   positive constant. This behavior is expected, because $ d=0 $ limit is a singular limit where the
   capacitive energy $ E_{c} $ vanish.

   We also perform the simulations of charge density distributions and capacitive energy in spherical
   geometry. The results are similar to those shown in Figs. 3-6 in disk geometry. 

 From the results in both disk and spherical geometries,
 we conclude that the quasi-hole energy diverges linearly as the system size grows
 which is much  higher than that of meron.

\section{ Conclusion }

  In this paper, we investigated the properties of  some known wavefunctions
  and their relations.  
  The existence of the gapless mode in the interlayer coherent Quantum Hall state
  dramatically change the properties of the wavefunctions of the
  ground state, quasi-hole and quasi-particle.
  We studied the relation between $ (111) $ wavefunction ( Eqn.\ref{111} ) and
  the EPQFM wavefunction ( Eqn.\ref{ex} ) and found that they are equivalent. We also
  investigated the properties of meron wavefunction ( Eqn.\ref{meron} ) and the quasi-hole wave
  function ( see Eqn.\ref{qh} ). These two wavefunctions differ only in normalization factors.
   Although normalizations factors are presumably not important in SLQH systems ( see section IIA)
  because of the gap in the bulk, we find that the charge distributions and
  energies of trial wavefunctions in BLQH are very sensitive to the normalization factors.
  This is because there is a neutral gapless mode even in the bulk in BLQH.
  We find that although the smallest meron has a
  well localized charge $ 1/2 $ and logarithmically divergent energy, 
  the charge of the quasi-hole excitation extends
  over the whole system and its energy diverges linearly with $  N $.
  This is because the gapless mode dramatically suppresses the relative density
  fluctuations between the two layers at any finite separations.
  These results indicate that the quasi-hole wavefunction is not a good
  trial wavefunction for the excitations of $ (111) $ state.

  As pointed out previously
  in Ref.\cite{moon}, $ (111) $ wavefunction may not be a qualitatively good wave function
  due to the existence of the gapless mode of the easy-plane ferromagnet at finite $ d $.
  Unfortunately, the qualitatively good ground state wavefunction is still unknown.
  It is still unknown if the energy of the quasi-hole wave
  function Eqn.\ref{qh} will become logarithmically if we replace the (111) wavefunction by
  the unknown qualitatively good ground state wavefunction.

   An important open question is to determine what is the lowest energy excitation:
  a pair of meron or conventional charge 1 excitation in Eqn.\ref{charge1} ? As shown in the text, the  energy of
  the naive candidate for the wavefunction of a pair of meron Eqn.\ref{pair} increases quadratically
  instead of logarithmically. Again, it is not known if the energy of the pair will
  become logarithmically if we replace the (111) wavefunction by
  the unknown qualitatively good ground state wavefunction.
  Without knowing a qualitatively good wavefunction for a pair
  of merons, it is hard to answer this open question precisely.

  In general, trial wave function approach is a very robust
  approach to study SLQH and multi-components systems as long as
  there is a gap in the bulk. The gap protects the properties of the system such as charge density
  distributions and energies from being sensitive to some subtle details of wavefunctions.
  However, the results of this paper indicate that
  qualitatively good trial wave functions for the interlayer coherent
  bilayer quantum Hall system at finite $ d $, especially for the excitations, are still
  unknown and searching for them remains an important open problem.
  Fortunately, 
  effective theory approaches such as EPQFM approach \cite{yang,moon,rev}  and  Composite Boson
  theory approach \cite{jap,moon,mcf,tri,prl} are  very effective
  to bring out most of the interesting phenomena in this system. In fact, all these effective theories
  start from the insights gained from Halperin's $ (111) $ wavefunction. Although many interesting
  results can be achieved qualitatively from effective theories without knowing the precise wavefunctions
  for the ground state and excitations, some quantitative results such as the lowest energy gap,
  quantum phase transition at $ d_{c} $, finite temperature phase transition, etc can be
  better achieved if we have a better knowledge of  the microscopic wavefunctions.

    This work is partially supported by NSF under grants Nos. DMR-0240458 ( GSJ ).
    We thank J. K. Jain and A. H. Macdonald for many insightful discussions and
    communications throughout this work.


\begin{references}
\bibitem{mis} There is a misprint in \cite{rev} where the charge of the quasi-particle was identified
              as 1 instead of $ 1/2 $.
\end{references}
\end{document}